\documentclass[letterpaper, 10pt, conference]{ieeeconf}

\IEEEoverridecommandlockouts
\usepackage{amsmath}
\usepackage{amssymb}
\usepackage{amsfonts}

\usepackage{graphicx}
\usepackage{booktabs}
\usepackage{multirow}
\usepackage{url}

\newtheorem{theorem}{Theorem}
\newtheorem{lemma}{Lemma}

\newtheorem{remark}{Remark}
\newtheorem{example}{Example}

\title{\LARGE \bf
A Matrix-Separation Approach for Stability of Discrete-Time Systems with Time-Varying Delay
}

\author{Nabil ECHAKROUNE, El Houssaine TISSIR, and Sadek BELAMFEDEL ALAOUI%
\thanks{N. ECHAKROUNE and E. H. TISSIR are with LESSI, Physics Dept., Faculty of Sciences Dhar El Mehraz, USMBA, Fez, Morocco (e-mails: {\tt\small nabil.echakroune@usmba.ac.ma}, {\tt\small elhoussaine.tissir@usmba.ac.ma}).}%
\thanks{S. BELAMFEDEL ALAOUI is with CLIC LAB, College of Computing, UM6P, Benguerir, Morocco (e-mail: {\tt\small sadek.belamfedel@um6p.ma}).}%
}

\begin{document}

\maketitle
\thispagestyle{empty}
\pagestyle{empty}
\begin{abstract}
This letter addresses the stability analysis of discrete-time linear systems with time-varying delay using a Lyapunov–Krasovskii functional (LKF) framework. An improved matrix-separation-based inequality (IMSBI) is developed to obtain tighter bounds for the summation terms arising in the forward difference of the LKF. Based on this result, an augmented LKF incorporating additional cross-terms is constructed to capture the interaction between system states and delay evolution. This leads to a less conservative delay-dependent stability criterion expressed as linear matrix inequalities (LMIs). Numerical results demonstrate improved stability margins and a reduced number of decision variables compared with existing methods.
\end{abstract}

\begin{keywords}
    Discrete-time systems, Augmented Lyapunov functional, Improved matrix-separation-based inequality, Stability analysis, Time-varying delay
\end{keywords}
\section{Introduction}

Discrete-time systems (DTS) form the backbone of modern digital control and arise across a wide range of engineering applications, including networked control,
	cyber-physical systems, and power systems~\cite{van2012discrete,yan2022lyapunov,zadeh2016discrete}. In such settings, delays due to communication latency, computational overhead, or actuator response times can degrade performance or cause instability~\cite{gu2003stability}. This motivates the study of DTS with time-varying delays (TVD), where a central problem is determining the largest delay range over which stability is preserved, and this has attracted sustained attention~\cite{chen2020stability,ge2022delay,zhang2022novel,zhang2023stabilitya,zhang2023stabilityb,wang2024delay,xie2024stability,luo2025new}.

The Lyapunov--Krasovskii functional (LKF) method, combined with linear matrix inequality (LMI) techniques, remains the dominant framework for analysing the stability of systems with delays~\cite{seuret2015hierarchy}. Reducing conservatism requires both a sufficiently rich LKF and tight inequality to bound the quadratic summation terms in its forward difference. Enriching the functional with finite differences of the state captures additional delay-state interactions~\cite{nabil2024improved}. Specifically, introducing a double-summation term involving the state, its first-order difference, and its second-order difference significantly enhances this capability. Regarding the bounding aspect, Jensen inequality~\cite{jiang2005stability} were progressively tightened by Wirtinger-based inequality~\cite{nam2015discretea}, auxiliary-function-based inequality~\cite{nam2015discreteb}, and other extensions, complemented by reciprocally convex lemmas~\cite{park2011reciprocally,zhang2017stability} for handling delay-partitioned terms. This work focuses on the key challenge of bounding the quadratic summation that arises from the forward difference of the LKF.

	Two main strategies exist for bounding this term. The first
	applies Jensen- or Wirtinger-type inequalities treating the
	state and its first-order difference as a single augmented
	vector~\cite{kwon2013stability,park2016stability,park2017stability}.
	The second uses matrix-separation-based inequalities (MSBI),
	which partition the weighting matrix to separately exploit the
	state and its first-order difference~\cite{chen2022delay,zhang2023stabilityb,sha2024stability}. In~\cite{zhang2023stabilityb} parameter-dependent matrices tighten the bound. However, handling the resulting reciprocally convex terms through the zero-equation method increases the number of decision variables while yielding only limited reductions in conservatism. 
    Reference~\cite{sha2024stability}
	reduces this count by combining MSBI with the reciprocally
	convex combination lemma, at the cost of a slight increase in
	conservatism. A common limitation of both strategies is the
	omission of the second-order state difference from the bounding
	procedure, leaving potentially useful information unexploited.

	Based on the above observations, this letter makes the following contributions:\vspace{-3pt}
	\begin{enumerate}
		\item[(i)] An improved MSBI (IMSBI, Lemma~\ref{lem3}) is
		derived that simultaneously accounts for the state, its
		first-order difference, and its second-order difference.
		\item[(ii)] An augmented LKF is constructed (Section~III)
		that incorporates first-order difference vectors absent
		in~\cite{nabil2024improved}, enabling the forward
		difference to exploit richer delay-state information.
		\item[(iii)] A delay-variation-independent stability
		criterion (Theorem~\ref{Theorem1}) is obtained, verified
		to be less conservative than state-of-the-art methods,
		including those that explicitly use delay-variation-rate
		information, while requiring $187n^2+16n$ decision
		variables --- best among state-of-the-art methods.
	\end{enumerate}
    \vspace{-2pt}
\textbf{Notation.}
$\mathbb{R}^{n}$, $\mathbb{R}^{m\times n}$, and $\mathbb{S}^{n}$
($\mathbb{S}^{n}_{+}$) denote the sets of $n$-dimensional real vectors,
$m{\times}n$ real matrices, and $n{\times}n$ symmetric (positive definite)
matrices, respectively.
For any square matrix $A$: $\mathrm{sym}\{A\}=A+A^{\top}$,
$\tilde{A}=\mathrm{diag}\{A,A\}$, and
$\Pi_i(A)=\mathrm{diag}\{A,3A,\ldots,(2i+1)A\}$ ($i+1$ blocks).
The shorthand $q_i(s)=s+i$, $i=1,2$, is used throughout.
For a vector-valued sequence $x(\kappa)$, define the first and
second-order differences $\zeta(\kappa)=x(\kappa+1)-x(\kappa)$ and
$\Delta\zeta(\kappa)=\zeta(\kappa+1)-\zeta(\kappa)$.
For integers $a>b\geq 0$, the single and double accumulation operators are:
$u_1(a,b)=\sum_{i=\kappa-a}^{\kappa-b} x(i),
	u_2(a,b)=\sum_{i=\kappa-a}^{\kappa-b}\sum_{j=i}^{\kappa-b} x(j),
	v_1(a,b)=\sum_{i=\kappa-a}^{\kappa-b} \zeta(i),
	v_2(a,b)=\sum_{i=\kappa-a}^{\kappa-b}\sum_{j=i}^{\kappa-b} \zeta(j)$.
\section{Preliminaries and problem formulation}
\vspace{-2pt}
Consider the following DTS, which includes a TVD
\vspace{-5pt}
	\begin{equation}\label{eq2}
		\left\{
		\begin{aligned}
			&x(\kappa+1)=Bx(\kappa)+B_{d}x(\kappa-d(\kappa))\\
			&x(\kappa)=\varphi(\kappa),\;\kappa\in\{-d_{u},\ldots,0\}
		\end{aligned}
		\right. 
	\vspace{-5pt}\end{equation}
    Here, $x(\kappa)\in\mathbb{R}^{n}$ represents the state vector, with the initial condition denoted as $\varphi(\kappa)$. The matrices $B$ and $B_{d}$ are real constant matrices with compatible dimensions. The TVD $d(\kappa):\mathbb{Z}_+\to\mathbb{Z}_+$ is an integer-valued function satisfying
	\vspace{-5pt}
		\begin{equation}\label{eq3}
			1\leq d_{l}\leq d(\kappa)\leq d_{u},\vspace{-5pt}
	\end{equation}
with known positive integer bounds $d_{l}$ and $d_{u}$. The objective of this letter is to establish a less conservative delay-dependent stability criterion for system \eqref{eq2} under the constraint \eqref{eq3}.\\
Before proceeding, we present some lemmas, as they are crucial for proving the theoretical contributions of the letter.	
\begin{lemma}[\cite{nam2015discreteb}]\label{lem1}
For a  matrix $R\in \mathbb{S}_{+}^{n}$, positive integers $\ell_{2}>\ell_{1}$, and vector-valued function $x(\kappa)$, The subsequent inequalities are valid:
\vspace{-5pt}
	\begin{align}
		&\sum_{i=\kappa-\ell_{2}}^{\kappa-\ell_{1}-1}\!x^{\top}(i)R x(i)\!\geq \!\frac{1}{\ell} \chi^{\top}(\kappa)\Omega^{\top}_{0}(\ell)\Pi_{1}(R)\Omega_{0}(\ell)\chi(\kappa),\label{eq4}\\
		&\sum_{i=\kappa-\ell_{2}}^{\kappa-\ell_{1}-1}\Delta x^{\top}(i)R\Delta x(i)\geq \frac{1}{\ell}\chi^{\top}(\kappa)\Omega^{\top}_{1}\Pi_{2}(R)\Omega_{1}\chi(\kappa),\label{eq5}\\
		&\sum_{i=\kappa-\ell_{2}}^{\kappa-\ell_{1}-1}\Delta\zeta^{\top}(i)R\Delta\zeta(i)\geq \frac{1}{\ell}\chi^{\top}(\kappa)\Omega^{\top}_{2}\Pi_{2}(R)\Omega_{2}\chi(\kappa),\label{eq6}
	\vspace{-5pt}\end{align}
	where $\ell=\ell_{2}-\ell_{1}$,
    \vspace{-3pt}
    \begin{align*}
		 \Omega_{0}(\ell)\!&=\!\mathrm{col}\left\lbrace q_{1}(\ell)\bar{e}_{3}-\bar{e}_{1},2q_{2}(\ell)\bar{e}_{4}-q_{1}(\ell)\bar{e}_{3}-\bar{e}_{1} \right\rbrace,\\[-1pt]
        \Omega_{1}\!&=\!\mathrm{col}\left\lbrace \bar{e}_{1}-\bar{e}_{2},\bar{e}_{1}+\bar{e}_{2}-2\bar{e}_{3},\bar{e}_{1}-\bar{e}_{2}+6\bar{e}_{3}-12\bar{e}_{4} \right\rbrace\!\!,\\[-1pt]
        \Omega_{2}\!&=\!\mathrm{col}\left\lbrace \bar{e}_{5}-\bar{e}_{6},\bar{e}_{5}+\bar{e}_{6}-2\bar{e}_{7},\bar{e}_{5}-\bar{e}_{6}+6\bar{e}_{7}-12\bar{e}_{8} \right\rbrace\!\!,\\[-1pt]
        \chi(\kappa)\!&=\!\mathrm{col}\left\lbrace x(\kappa-\ell_{1}),x(\kappa-\ell_{2}),\frac{u_{1}(\ell_{2},\ell_{1})}{q_{1}(\ell)},\frac{u_{2}(\ell_{2},\ell_{1})}{q_{1}(\ell)q_{2}(\ell)},\right.\\[-1pt]
       &\quad\quad\;\;\left.\zeta(\kappa-\ell_{1}),\zeta(\kappa-\ell_{2}),\frac{v_{1}(\ell_{2},\ell_{1})}{q_{1}(\ell)},\frac{v_{2}(\ell_{2},\ell_{1})}{q_{1}(\ell)q_{2}(\ell)}\right\rbrace,\\[-1pt]
      \bar{e}_{i}&=\left[\begin{array}{ccc} 0_{n,(i-1)n} & I_{n} & 0_{n,(8-i)n} \end{array}\right], i=1,2,...,8.
	\end{align*}
    \end{lemma}
    
\begin{lemma}[\cite{zhang2017stability}]\label{lem2} 
For any scalar $\alpha \in (0,1)$, matrices $R_1, R_2 \in \mathbb{S}^n_+$, and arbitrary matrices $X_1, X_2 \in \mathbb{R}^{n \times n}$, the following matrix inequality holds:
\vspace{-3pt}
\begin{equation}\label{eq7}
\begin{bmatrix}
\frac{1}{\alpha}R_1 & 0 \\
0 & \frac{1}{1-\alpha}R_2
\end{bmatrix} \!\succeq\!
\begin{bmatrix}
R_1 + (1-\alpha)r_1 & (1-\alpha)X_1 + \alpha X_2 \\
\ast & R_2 + \alpha r_2
\end{bmatrix}\!,\vspace{-5pt}
\end{equation}
where $r_1 = R_1 - X_2 R_2^{-1} X_2^\top$ and $r_2 = R_2 - X_1^\top R_1^{-1} X_1$.
\end{lemma}
\section{Main results}
\subsection{An Improved Matrix-Separation-Based Inequality}
This subsection introduces an improved matrix-separation-based inequality (IMSBI) aimed at enhancing the bounding accuracy of the summation term
\vspace{-5pt}
\begin{equation}\label{eq1}
    \mathrm{J}(x)=\sum_{i=\kappa-\ell_{2}}^{\kappa-\ell_{1}-1}\eta^{T}(i)R\eta(i),\vspace{-5pt}
\end{equation} 
where $\eta(i)=\mathrm{col}\left\lbrace x(i),\zeta(i),\Delta\zeta(i) \right\rbrace,$ and \( R \in \mathbb{S}_{+}^{3n} \).
\begin{lemma}\label{lem3} 
Let \( J(x) \) be defined as in \eqref{eq1}, where $R$ admits the decomposition:
\begin{align}\label{eq8}
\resizebox{0.89\hsize}{!}{$R \!\!=\!\!\!
\underbrace{\begin{bmatrix} R_{11} & R_{12}-T & R_{13} \\ \ast & R_{22}-T-Z & R_{23}-T \\ \ast & \ast & R_{33}-T-Z \end{bmatrix}}_{R_a} \!\!+\!\!
\underbrace{\begin{bmatrix} 0 & T & 0 \\ \ast & T & T \\ \ast & \ast & T \end{bmatrix}}_{R_b}\!\!+\!\!
\underbrace{\begin{bmatrix} 0 & 0 & 0 \\ \ast & Z & 0 \\ \ast & \ast & Z \end{bmatrix}}_{R_c}\!,$}
\end{align}
with \( R_{12}, R_{13}, R_{23} \in \mathbb{R}^{n \times n} \) and \( R_{11}, R_{22}, R_{33}, T, Z \in \mathbb{S}^n \). If \( R_a \succ 0 \) and \( Z \succ 0 \), then:
\vspace{-5pt}
\begin{equation}\label{eq9}
J(x) \geq \chi^{\top}(\kappa) \Psi_1(\ell) \chi(\kappa) + \frac{1}{\ell} \chi^{\top}(\kappa) F_1^{\top} \Psi_2 F_1 \chi(\kappa),\vspace{-5pt}
\end{equation}
where
\vspace{-3pt}
\begin{align*}
\Psi_1(\ell)\!&=\! \ell E_0^{\top} \Pi_1(R_a) E_0 + \mathrm{sym}\big\{ E_0^{\top} \Pi_1(R_a) E_1 \big\} + c_0^{\top} \mathcal{T} c_0, \\[-1pt]
\Psi_2 \!&= \!\sigma^{\top} \Pi_1(R_a) \sigma + \Pi_2(\tilde{Z}),  \\[-1pt]
E_0 \!&=\! \mathrm{col}\!\big\{ \bar{e}_3, \bar{e}_0, \bar{e}_0, 2\bar{e}_4 - \bar{e}_3, \bar{e}_0, \bar{e}_0 \big\}, \\[-1pt]
E_1\! &=\! \mathrm{col}\!\big\{ \bar{e}_3 - \bar{e}_1, \bar{e}_1 - \bar{e}_2, \bar{e}_5 - \bar{e}_6, 4\bar{e}_4 - \bar{e}_3 - \bar{e}_1, \nonumber \\[-1pt]
     &\qquad \bar{e}_1 + \bar{e}_2 - 2\bar{e}_3, \bar{e}_5 + \bar{e}_6 - 2\bar{e}_7 \big\},  \\[-1pt]
F_1 \!&=\! \mathrm{col}\!\big\{ \bar{e}_1 - \bar{e}_2, \bar{e}_5 - \bar{e}_6, \bar{e}_1 + \bar{e}_2 - 2\bar{e}_3, \bar{e}_5 + \bar{e}_6 - 2\bar{e}_7, \nonumber  \\[-1pt]
     &\qquad  \bar{e}_1 - \bar{e}_2 + 6\bar{e}_3 - 12\bar{e}_4, \bar{e}_5 - \bar{e}_6 + 6\bar{e}_7 - 12\bar{e}_8 \big\},\\[-1pt]
\sigma &= \begin{bmatrix}
-\frac{1}{2}I & 0 & -\frac{1}{2}I & 0 & 0 & 0 \\
I & 0 & 0 & 0 & 0 & 0 \\
0 & I & 0 & 0 & 0 & 0 \\
-\frac{1}{6}I & 0 & -\frac{1}{2}I & 0 & -\frac{1}{3}I & 0 \\
0 & 0 & I & 0 & 0 & 0 \\  
0 & 0 & 0 & I & 0 & 0
\end{bmatrix},\quad \bar{e}_{0}=0_{n\times 8n}, \\[-1pt]
\mathcal{T} &= \mathrm{diag}\left\{T,T,-T,-T\right\}, \quad 
c_0 = \mathrm{col}\left\{ \bar{e}_1, \bar{e}_5, \bar{e}_2, \bar{e}_6 \right\}.\vspace{-5pt}
\end{align*}
Additional notations, including $\chi(\kappa)$ and $\bar{e}_{i}$, are introduced and defined in Lemma~\ref{lem1}.
\end{lemma}
\begin{proof}
     Using the separation matrix \eqref{eq8} of $R$, $\mathrm{J}(x)$ can be expressed as the sum of three components:
\vspace{-5pt}
\begin{align}\label{eq10}
\resizebox{0.89\hsize}{!}{$\mathrm{J}(x) \!\!=\!\! \!
\underbrace{
\sum_{i=\kappa-\ell_{2}}^{\kappa-\ell_{1}-1} \!\!\!\!\eta^{\top}(i) R_{a} \eta(i)
}_{\mathrm{J}_{a}}\!+\!\!\!
\underbrace{
\sum_{i=\kappa-\ell_{2}}^{\kappa-\ell_{1}-1} \!\!\!\!\eta^{\top}(i) R_{b} \eta(i)
}_{\mathrm{J}_{b}} \!+\!\!\!
\underbrace{
\sum_{i=\kappa-\ell_{2}}^{\kappa-\ell_{1}-1} \!\!\!\!\eta^{\top}(i) R_{c} \eta(i)
}_{\mathrm{J}_{c}},$}
\end{align}
From Lemma~\ref{lem1}, it follows that $R_{a} > 0$ implies:
\vspace{-5pt}
\begin{align}
J_a &\geqslant \frac{1}{\ell} \chi^{\top}(\kappa) [E_1 + \ell E_0]^{\top} \Pi_1(R_a) [E_1 + \ell E_0] \chi(\kappa), \nonumber \\[-1pt]
&= \chi^{\top}(\kappa) \left[ \tfrac{1}{\ell} E_1^{\top} \Pi_1(R_a) E_1 + \mathrm{sym}\{E_0^{\top} \Pi_1(R_a) E_1\}\right.\nonumber\\[-1pt]
&\left.\qquad\qquad+ \ell E_0^{\top} \Pi_1(R_a) E_0 \right] \chi(\kappa). \label{eq11}\vspace{-5pt}
\end{align}
For $R_b$ and $T \in \mathbb{S}^n$, direct computation gives:
\vspace{-5pt}\begin{align}
J_b &= \sum_{i=\kappa-\ell_{2}}^{\kappa-\ell_{1}-1}\big[\zeta^{\top}(i)T\zeta(i)+2\zeta^{\top}(i)Tx(i)\nonumber\\[-1pt]
&\quad+\Delta\zeta^{\top}(i)T\Delta\zeta(i)+2\Delta\zeta^{\top}(i)T\zeta(i)\big],\nonumber\\[-1pt]
&=x^{\top}(\kappa-\ell_1)Tx(\kappa-\ell_1)-x^{\top}(\kappa-\ell_2)Tx(\kappa-\ell_2)\nonumber\\[-1pt]
&\quad+\zeta^{\top}(\kappa-\ell_1)T\zeta(\kappa-\ell_1)-\zeta^{\top}(\kappa-\ell_2)T\zeta(\kappa-\ell_2),\nonumber\\[-1pt]
&=\chi^{\top}(\kappa)\; c_0^{\top} \mathcal{T} c_0\; \chi(\kappa).\label{eq12}\vspace{-5pt}
\end{align}
With $Z \succ 0$ and via \eqref{eq5},\eqref{eq6}, we bound $J_c$:
\vspace{-5pt}
\begin{equation}\label{eq13}
J_c \geqslant \frac{1}{\ell} \chi^{\top}(\kappa) F_1^{\top} \Pi_2(\tilde{Z}) F_1 \chi(\kappa)
\end{equation}
Combining \eqref{eq10}, \eqref{eq11}, \eqref{eq12}, \eqref{eq13} with the relation $E_1 = \sigma F_1$ establishes \eqref{eq9}. Hence, the proof of Lemma~\ref{lem3} is concluded.
\end{proof}

\begin{remark}\label{rem1}
Lemma~\ref{lem3} generalizes the MSBI in \cite{zhang2023stabilityb}: 
omitting $\Delta\zeta(i)$ from $\eta(i)$ and substituting $T=T_1+\ell T_2$, 
$Z=Z_1+\ell Z_2$, and further considering the zero-equation 
$2\Big(\frac{F_1}{\ell}\chi(\kappa)-\frac{F_1}{\ell}\chi(\kappa)\Big)^{\top}\ell M^{\top}\chi(\kappa)=0$, yields its $m=1$ case. However, handling the reciprocally convex 
terms in~\eqref{eq9} via this technique, see Lemma~6 in 
\cite{nabil2024improved}, requires free matrices 
$M_1,M_2\in\mathbb{R}^{8n\times 6n}$, introducing $96n^2$ decision 
variables per application of the IMSBI. By contrast, combining 
Lemma~\ref{lem2} with~\eqref{eq9} uses only 
$X_1,X_2\in\mathbb{R}^{6n\times 6n}$, reducing this count to $72n^2$, a 
$25\%$ saving at the cost of a small increase in conservatism. Since the 
free-matrix dimensions in \cite{zhang2023stabilityb} grow with the order 
parameter $m$, any selection $m>1$ incurs substantially more decision 
variables than the proposed approach.
\end{remark}

\begin{remark}\label{rem2}
Existing methods estimate $J(\kappa)$ with limited inter-component exploitation:\cite{kwon2013stability,park2016stability,park2017stability} 
use Jensen/Wirtinger inequalities on $\mathrm{col}\{x(i),\zeta(i)\}$ without 
capturing component interactions with $\Delta \zeta$; the MSBI in \cite{zhang2023stabilityb} 
partially addresses this via matrix separation but remains restricted to 
$\{x(i),\zeta(i)\}$; and \cite{nabil2024improved} employs the full vector 
$\eta(i)=\mathrm{col}\{x(i),\zeta(i),\Delta\zeta(i)\}$ yet treats it 
monolithically. The proposed IMSBI overcomes these limitations by explicitly 
separating $T$ and $Z$ from $R$, thereby incorporating the pairwise 
relationships among $x(i)$, $\zeta(i)$, and $\Delta\zeta(i)$. In 
particular,~\eqref{eq12} yields a zero-error bound for the $T$-dependent 
subterm, while the boundary terms $\zeta(\kappa-\ell_1)$, 
$\zeta(\kappa-\ell_2)$ and the summation terms $v_1(\ell_2,\ell_1)$, 
$v_2(\ell_2,\ell_1)$ in~\eqref{eq9} supply additional delay-dependent 
information, jointly producing a tighter estimate of $J(\kappa)$.
\end{remark}
\subsection{Delay-Dependent Stability Criterion}
This subsection presents a new sufficient condition to ensure the asymptotic stability of DTS \eqref{eq2} in the presence of a TVD satisfying \eqref{eq3}. For brevity, the following notations are adopted:
   $ d_{21}= d_{u} - d_{l},\! 
    \quad\!
    d_{\kappa1} = d(\kappa) - d_{l},\! 
    \quad\!
    d_{\kappa2} = d_{u} - d(\kappa)$.
\begin{theorem}\label{Theorem1} Given positive integers $d_{l}\geq 1$ and $d_{u}>d_{l}$, system~\eqref{eq2} subject to~\eqref{eq3} is asymptotically stable if there exist matrices $P_{1}\in\mathbb{S}^{9n}_{+}, P_{2}\in\mathbb{S}^{5n}_{+}, Q_{i}\in\mathbb{S}^{2n}_{+}(i=1,2,3), R=\left[R_{ij}\right]\in\mathbb{S}^{3n}_{+}, S=\left[S_{ij}\right]\in\mathbb{S}^{3n}_{+}, Z_{i}\in\mathbb{S}^{n}_{+}(i=1,2,3)$, and $T_{i}\in\mathbb{S}^{n}(i=1,2,3)$, along with matrices $X_{1}, X_{2}, G_{i}(i=1,2,3,4)$, such that the subsequent inequalities remain true for each $d(\kappa)\in\left\lbrace d_{l},d_{u}\right\rbrace$:
	\begin{align}\label{eq14}
		S_{a}\succ0, \quad R_{ia}\succ0,\quad \Psi_{2i}\succ0 \quad i=1,2,
        \end{align}\vspace{-10pt}
        \begin{align}\label{eq15}
	   \begin{bmatrix}
			\Xi(d(\kappa)) & \frac{d_{\kappa1}}{d_{21}}F^{\top}_{12}X^{\top}_{1}+\frac{d_{\kappa2}}{d_{21}}F^{\top}_{11}X_{2} \\
			\ast & -d_{\kappa1}\Psi_{21}-d_{\kappa2}\Psi_{22}
		\end{bmatrix} \prec0,
	\end{align}
	where:
   \vspace{-3pt}
		\begin{align*}
		\Xi(d(\kappa))&=\Phi(d(\kappa))-\Psi(d(\kappa)),\\[-1.5pt]
        \Psi_{2i}&=\sigma^{\top}\Pi_{1}(R_{ia})\sigma+\Pi_{2}(\tilde{Z}_{i}),\quad i=1,2,\\[-1.3pt]
		\Phi(d(\kappa))&=\Phi_{1}(d(\kappa))+\Phi_{2}+\Phi_{3}+\Phi_{4}(d(\kappa)),\\[-1.3pt]
		\Psi(d(\kappa))&=\Psi_{11}(d(\kappa))+\Psi_{12}(d(\kappa))+\Psi_{3}+\Psi_{4}(d(\kappa)),\\[-1.3pt]
		\Phi_{1}(d(\kappa))&=\Upsilon^{\top}_{1}(d(\kappa))P_{1}\Upsilon_{1}(d(\kappa))+\Upsilon^{\top}_{2}(d(\kappa))P_{2}\Upsilon_{2}(d(\kappa))\\[-1.3pt]
        &\quad-\!\left[\!\Gamma^{\top}_{1}\!(d(\kappa))P_{1}\Gamma_{1}\! (d(\kappa))\!+\!\Gamma^{\top}_{2}\!(d(\kappa))P_{2}\Gamma_{2}(d(\kappa))\right]\!\!,\\[-1.3pt]
	    \Phi_{2}&= c_{1}^{\top}\left[ Q_{1}+q_{1}(d_{21})Q_{3}\right]c_{1}+c_{2}^{\top}\left[ Q_{2}-Q_{1}\right]c_{2}\\[-1.3pt]
        &\quad-c_{3}^{\top}Q_{3}c_{3}-c_{4}^{\top}Q_{2}c_{4},\\[-1.3pt]
        \Phi_{3}&= c_{5}^{\top}\left[ d_{21}R+d_{l}S\right]c_{5},
        \end{align*}
        \begin{align*}\vspace{-18mm}
        \Phi_{4}(d(\kappa))&=\mathrm{sym}\Big\{ \sum_{i=1}^{4}\Phi_{4i}(d(\kappa))\Big\},\\[-1pt]
        \Phi_{41}(d(\kappa))&=e_{g}^{\top}G_{1}\big[e_{12}+e_{2}-e_{3}-q_{1}(d_{\kappa1})e_{15}\big],\\[-1pt]
        \Phi_{42}(d(\kappa))&=e_{g}^{\top}G_{2}\big[e_{13}+e_{3}-e_{4}-q_{1}(d_{\kappa2})e_{16}\big],\\[-1pt]
        \Phi_{43}(d(\kappa))&=e_{g}^{\top}G_{3}\big[e_{12}+e_{2}-e_{6}-q_{2}(d_{\kappa1})e_{17}\big],\\[-1pt]
        \Phi_{44}(d(\kappa))&=e_{g}^{\top}G_{4}\big[e_{13}+e_{3}-e_{7}-q_{2}(d_{\kappa2})e_{18}\big],\\[-1pt]
        \Psi_{1i}(d(\kappa))&=d_{\kappa i}E^{\top}_{0i}\Pi_{1}(R_{ia})E_{0i}+\mathrm{sym}\big\{E^{\top}_{0i}\Pi_{1}(R_{ia})E_{1i}\big\}\\[-1pt]
        &\quad+c^{\top}_{0i}\mathcal{T}_{i}c_{0i},\quad i=1,2,\\[-1pt]
        \Psi_{3}&=\frac{1}{d_{l}}\big[F^{\top}_{2}\Pi_{1}(S_{a})F_{2}+F^{\top}_{3}\Pi_{2}(\tilde{Z}_{3})F_{3}\big]\\[-1pt]
        &\quad+c^{\top}_{03}\mathcal{T}_{3}c_{03},\\[-1pt]
    \Psi_{4}(d(\kappa)) &=
\frac{1}{d_{21}}\begin{bmatrix} F_{11} \\ F_{12} \end{bmatrix}^\top
\Psi_{2}\begin{bmatrix} F_{11} \\ F_{12} \end{bmatrix},\\[-1pt]
        \Gamma_{1}(d(\kappa))\!&=\!\mathrm{col}\big\{ e_{11},e_{1},e_{2},e_{3},e_{4},q_{1}(d_{l})e_{5}-e_{1},q_{1}(d_{\kappa1})e_{6} \\[-1pt]
        &\quad \!\!+q_{1}(d_{\kappa2})e_{7}-e_{2}-e_{3},q_{1}(d_{l})q_{2}(d_{l})e_{8} \\[-1pt]
        &\quad\!\! -q_{1}(d_{l})e_{1},q_{1}(d_{\kappa1})q_{2}(d_{\kappa1})e_{9}+q_{1}(d_{\kappa2})q_{2}(d_{\kappa2}) \\[-1pt]
        &\quad e_{10}+d_{\kappa2}q_{1}(d_{\kappa1})e_{6}-q_{1}(d_{21})e_{2}-q_{1}(d_{\kappa2})e_{3}\big\},\\[-1pt]
        \Gamma_{2}(d(\kappa))\!&=\!\mathrm{col}\big\{ e_{1}\!-\!e_{11},e_{1}-e_{2},q_{1}(d_{\kappa1})e_{15}+q_{1}(d_{\kappa2})e_{16}\\[-1pt]
        &\quad\!\!-e_{12}\!-\!e_{13},q_{1}(d_{l})(e_{1}-e_{5}),q_{1}(d_{\kappa1})q_{2}(d_{\kappa1})e_{17}\\[-1pt]
        &\quad\!\! +q_{1}(d_{\kappa2})q_{2}(d_{\kappa2})e_{18}+d_{\kappa2}q_{1}(d_{\kappa1})e_{15} \\[-1pt]
        &\quad\!\! -q_{1}(d_{21})e_{12}-q_{1}(d_{\kappa2})e_{13}\big\},\\[-1pt]
        \Upsilon_{1}(d(\kappa))\!&=\!\mathrm{col}\big\{ e_{1},e_{s},e_{2}+e_{12},e_{3}+e_{13},e_{4}+e_{14},q_{1}(d_{l}) \\[-1pt]
        &\quad e_{5}-e_{2},q_{1}(d_{\kappa1})e_{6}+q_{1}(d_{\kappa2})e_{7}-e_{3} -e_{4},\\[-1pt]
        &\quad q_{1}(d_{l})q_{2}(d_{l})e_{8}-q_{1}(d_{l})e_{5},q_{1}(d_{\kappa1})q_{2}(d_{\kappa1})e_{9}\\[-1pt]
        &\quad\! +q_{1}(d_{\kappa2})q_{2}(d_{\kappa2}) e_{10}+(d_{\kappa2}-1)q_{1}(d_{\kappa1})e_{6} \\[-1pt]
        &\quad \!-q_{1}(d_{\kappa2})e_{7}-d_{\kappa2}e_{3}\big\},\\[-1pt]
        \Upsilon_{2}(d(\kappa))\!&=\!\mathrm{col}\big\{ e_{s}-e_{1},e_{s}-e_{2}-e_{12},q_{1}(d_{\kappa1})e_{15}\\[-1pt]
        &\quad\!\! +q_{1}(d_{\kappa2})e_{16}-e_{13}-e_{14},d_{l}e_{s}+e_{2}-q_{1}(d_{l}) \\[-1pt]
        &\quad\!\! e_{5},q_{1}(d_{\kappa1})q_{2}(d_{\kappa1})e_{17}+q_{1}(d_{\kappa2})q_{2}(d_{\kappa2})e_{18} \\[-1pt]
        &\quad\!\!+\!(d_{\kappa2}\!-\!1)q_{1}(d_{\kappa1})e_{15}-q_{1}(d_{\kappa2})e_{16}-d_{\kappa2}e_{13}\big\},\\[-1pt]
        E_{0i}&=\mathrm{col}\big\{ e_{5+i},e_{0},e_{0},2e_{8+i}-e_{5+i},e_{0},e_{0} \big\},i\!=\!1,2,\\[-1pt]
        E_{1i}&=\mathrm{col}\big\{ e_{5+i}-e_{1+i},e_{1+i}-e_{2+i},e_{11+i}-e_{12+i}, \\[-1pt]
        &\quad 4e_{8+i}-e_{5+i}-e_{1+i},e_{1+i}+e_{2+i}-2e_{5+i}, \\[-1pt]
        &\quad e_{11+i}+e_{12+i}-2e_{14+i}\big\},\;i=1,2,\\[-1pt]
        F_{1i}&=\mathrm{col}\big\{ e_{1+i}-e_{2+i},e_{11+i}-e_{12+i},e_{1+i}+e_{2+i} \\[-1pt]
        &\quad -2e_{5+i},e_{11+i}+e_{12+i}-2e_{14+i},e_{1+i}-e_{2+i} \\[-1pt]
        &\quad +6e_{5+i}-12e_{8+i},e_{11+i}-e_{12+i}+6e_{14+i}\\[-1pt]
        &\quad -12e_{16+i}\big\},\;i=1,2,\\[-1pt]
        F_{2}&=\mathrm{col}\big\{ q_{1}(d_{l})e_{5}-e_{1},e_{1}-e_{2},e_{s}-e_{1}-e_{12}, \\[-1pt]
        &\quad 2q_{2}(d_{l})e_{8}-q_{1}(d_{l})e_{5}-e_{1},e_{1}+e_{2}-2e_{5}, \\[-1pt]
        &\quad \frac{d_{l}-1}{q_{1}(d_{l})}e_{s}-e_{1}+e_{12}+\frac{2}{q_{1}(d_{l})}e_{2}\big\},\\[-1pt]
        F_{3}&=\mathrm{col}\big\{ e_{1}-e_{2},e_{s}-e_{1}-e_{12},e_{1}+e_{2}-2e_{5}, \\[-1pt]
        &\quad \frac{d_{l}-1}{q_{1}(d_{l})}e_{s}-e_{1}+e_{12}+\frac{2}{q_{1}(d_{l})}e_{2},e_{1}-e_{2}+6e_{5} \\[-1pt]
        &\quad -12e_{8},\frac{(d_{l}-1)(d_{l}-2)}{q_{1}(d_{l})q_{2}(d_{l})}e_{s}-e_{1}-e_{12}\\[-1pt]
        &\quad-\frac{6}{q_{1}(d_{l})}e_{2}+\frac{12}{q_{2}(d_{l})}e_{5}\big\},
         \end{align*}
        \begin{align*}
\Psi_{2}&=\begin{bmatrix}
(2-\tfrac{d_{\kappa1}}{d_{21}})\Psi_{21} & (1-\tfrac{d_{\kappa1}}{d_{21}})X_{1} + \tfrac{d_{\kappa1}}{d_{21}}X_{2} \\
\ast & (1+\tfrac{d_{\kappa1}}{d_{21}})\Psi_{22}
\end{bmatrix},\\[-1pt]
R_{ia}&=\begin{bmatrix} R_{11} & R_{12}-T_{i} & R_{13}\\ \ast & R_{22}-T_{i}-Z_{i} & R_{23}-T_{i} \\ \ast & \ast & R_{33}-T_{i}-Z_{i}\end{bmatrix}\!,\;i=1,2\\[-1pt]
        S_{a}&=\begin{bmatrix} S_{11} & S_{12}-T_{3} & S_{13}\\ \ast & S_{22}-T_{3}-Z_{3} & S_{23}-T_{3} \\ \ast & \ast & S_{33}-T_{3}-Z_{3}\end{bmatrix},\\[-1pt]
\mathcal{T}_{i}&=\mathrm{diag}\big\{ T_{i},T_{i},-T_{i},-T_{i}\big\} \;i=1,2,3,\\[-1pt]
c_{0i}&=\mathrm{col}\big\{ e_{1+i},e_{11+i},e_{2+i},e_{12+i}\big\},\;i=1,2,\\[-1pt]
       c_{03}&=\mathrm{col}\big\{ e_{1},e_{s}-e_{1},e_{2},e_{12}\big\},\quad c_{1}=\mathrm{col}\big\{ e_{1},e_{s}-e_{1}\big\},\\[-1pt]
        c_{i}&=\mathrm{col}\big\{ e_{i},e_{10+i}\big\}, i=2,3,4,\quad e_{s}=Be_{1}+B_{d}e_{3},\\[-1pt]
        c_{5}&=\mathrm{col}\big\{ e_{11},e_{1}-e_{11},e_{s}-2e_{1}+e_{11}\big\},\\[-1pt] 
        e_{g}&=\begin{bmatrix} e^{\top}_{2}, e^{\top}_{3}, e^{\top}_{4}, e^{\top}_{6}, e^{\top}_{7}, e^{\top}_{12}, e^{\top}_{13}, e^{\top}_{15}, e^{\top}_{16}, e^{\top}_{17}, e^{\top}_{18}\end{bmatrix}^{\top},\\[-1pt]
        e_{i}&=\begin{bmatrix} 0_{n,(i-1)n}, I_{n}, 0_{n,(18-i)n} \end{bmatrix}^{\top}, i=1,2,...,18.
	\end{align*}
 \end{theorem}
 
\begin{proof}
    To prove asymptotic stability, we construct a candidate LKF of the form,
    \vspace{-8pt}
\begin{equation}\label{eq16}
		V(\kappa)=\sum_{i=1}^{3}V_{i}(\kappa),\vspace{-8pt}
\end{equation}
 where
   \vspace{-5pt}
	\begin{align*}
			V_{1}(\kappa)&=\eta_{11}^{\top}(\kappa)P_{1}\eta_{11}(\kappa)+\eta_{12}^{\top}(\kappa)P_{2}\eta_{12}(\kappa),\\[-1pt]
			V_{2}(\kappa)&=\sum_{i=\kappa-d_{l}}^{\kappa-1}\eta_{2}^{\top}(i)Q_{1}\eta_{2}(i)+\sum_{i=\kappa-d_{u}}^{\kappa-d_{l}-1}
			\eta_{2}^{\top}(i)Q_{2}\eta_{2}(i)\\[-1pt]
            &\quad+\sum_{j=-d_{u}}^{-d_{l}}\sum_{i=\kappa+j}^{\kappa-1}\eta_{2}^{\top}(i)Q_{3}\eta_{2}(i),\\[-1pt]
			V_{3}(\kappa)&=\sum_{j=-d_{u}}^{-d_{l}-1}\sum_{i=\kappa+j}^{\kappa-2}\eta_{3} ^{\top}(i)R\eta_{3}(i)\\[-1pt]
			&\quad+\sum_{j=-d_{l}}^{-1}\sum_{i=\kappa+j}^{\kappa-2}\eta_{3} ^{\top}(i) S \eta_{3}(i),\\[-1pt]
		\eta_{11}(\kappa)&=\mathrm{col}\left\lbrace x(\kappa-1),x(\kappa),x(\kappa-d_{l}),x(\kappa-d(\kappa)),\right.\\[-1pt]
        &\quad\left.x(\kappa-d_{u}),
		u_{1}(d_{l},1),u_{1}(d_{u},d_{l}+1),
		u_{2}(d_{l},1),\right.\\
        &\quad\left.u_{2}(d_{u},d_{l}+1)\right\rbrace,\\[-1pt]
        \eta_{12}(\kappa)&=\mathrm{col}\left\lbrace \zeta(\kappa-1),v_{1}(d_{l},1),v_{1}(d_{u},d_{l}+1),
		v_{2}(d_{l},1),\right.\\[-1pt]
        &\quad\quad\left.v_{2}(d_{u},d_{l}+1)\right\rbrace,\\[-1pt]
		\eta_{2}(i)&= \mathrm{col}\left\lbrace x(i),\zeta(i)\right\rbrace,\!\quad\! \eta_{3}(i)= \mathrm{col}\left\lbrace x(i),\zeta(i),\Delta\zeta(i)\right\rbrace,\vspace{-5pt}
	\end{align*}
	We define the augmented state vector $\xi(\kappa)$ as follows:
  \vspace{-5pt}
	\begin{align*}
\xi(\kappa)\!&=\! \mathrm{col}\big\{x(\kappa),x(\kappa-d_{l}),x(\kappa-d(\kappa)),x(\kappa-d_{u}),\\
&\quad\!\frac{u_{1}(d_{l},0)}{q_{1}(d_{l})},\frac{u_{1}(d(\kappa),d_{l})}{q_{1}(d_{\kappa1})},\frac{u_{1}(d_{u},d(\kappa))}{q_{1}(d_{\kappa2})},\frac{u_{2}(d_{l},0)}{q_{1}(d_{l})q_{2}(d_{l})},\\
&\quad \!\frac{u_{2}(d(\kappa),d_{l})}{q_{1}(d_{\kappa1})q_{2}(d_{\kappa1})},\frac{u_{2}(d_{u},d(\kappa))}{q_{1}(d_{\kappa2})q_{2}(d_{\kappa2})},x(\kappa-1),\\
    &\quad\!\zeta(\kappa-d_{l}),\zeta(\kappa-d(\kappa)),\zeta(\kappa-d_{u}),\frac{v_{1}(d(\kappa),d_{l})}{q_{1}(d_{\kappa1})},\\
    &\quad\!\frac{v_{1}(d_{u},d(\kappa))}{q_{1}(d_{\kappa2})},\frac{v_{2}(d(\kappa),d_{l})}{q_{1}(d_{\kappa1})q_{2}(d_{\kappa1})},\frac{v_{2}(d_{u},d(\kappa))}{q_{1}(d_{\kappa2})q_{2}(d_{\kappa2})}\big\}.\vspace{-5pt}
	\end{align*}
	It can be verified that the LKF vectors satisfy the relations:
     \vspace{-5pt}
	\begin{align*}
	\eta_{11}(\kappa)&= \Gamma_{1}(d(\kappa))\xi(\kappa),\quad \eta_{11}(\kappa+1)= \Upsilon_{1}(d(\kappa))\xi(\kappa),\\
    \eta_{12}(\kappa)&= \Gamma_{2}(d(\kappa))\xi(\kappa),
    \quad \eta_{12}(\kappa+1)= \Upsilon_{2}(d(\kappa))\xi(\kappa).
	\end{align*}
    Since $P_1\in\mathbb{S}^{9n}_+$, $P_2\in\mathbb{S}^{5n}_+$, 
$Q_i\in\mathbb{S}^{2n}_+$, and $R,S\in\mathbb{S}^{3n}_+$, 
it follows directly that $V(\kappa)>0$ for all $\xi(\kappa)\neq 0$. Next, by computing the forward difference of the functional  $V(\kappa)$ along the solutions of system \eqref{eq2}, we derive:
    \vspace{-5pt}
	\begin{align}
		\Delta V_{1}(\kappa)\!&=\! \eta^{\top}_{11}(\kappa\!+\!1)P_{1}\eta_{11}(\kappa\!+\!1)+\eta^{\top}_{12}(\kappa\!+\!1)P_{2}\eta_{12} 
             (\kappa\!+\!1)\nonumber\\
            &\quad\!-\left[\eta^{\top}_{11}(\kappa)P_{1}\eta_{11}(\kappa)+\eta^{\top}_{12}(\kappa)P_{2}\eta_{12}(\kappa)\right],\nonumber\\
			\!&=\!  \xi^{\top}(\kappa)\Phi_{1}(d(\kappa))\xi(\kappa),\label{eq17}\\[+3pt]
		\Delta V_{2}(\kappa)\!&=\!\eta^{\top}_{2}(\kappa)Q_{1}\eta_{2}(\kappa)-\eta^{\top}_{2}(\kappa-d_{l})Q_{1}\eta_{2}(\kappa-d_{l})
			 \nonumber\\
    \!&\!\quad+\eta^{\top}_{2}(\kappa-d_{l})Q_{2}\eta_{2}(\kappa-d_{l})-\eta^{\top}_{2}(\kappa-d_{u})Q_{2}
			\nonumber\\
    \!&\!\quad\eta_{2}(\kappa-d_{u})+q_{1}(d_{21})\eta^{\top}_{2}(\kappa)Q_{3}\eta_{2}(\kappa)\nonumber\\
    \!&\!\quad-\sum_{i=\kappa-d_{u}}^{\kappa-d_{l}}\eta^{\top}_{2}(i)Q_{3}\eta_{2}(i),\nonumber\\
	\!&\!\leq  \eta^{\top}_{2}(\kappa)\left[Q_{1}+q_{1}(d_{21})Q_{3}\right] \eta_{2}(\kappa)+\eta^{\top}_{2}(\kappa- 
             d_{l})\left[Q_{2}\right.\nonumber\\
    \!&\!\quad\left.-Q_{1}\right]\eta_{2}(\kappa-d_{l})-\eta^{\top}_{2}(\kappa-d_{u})Q_{2}\eta_{2}(\kappa-d_{u})\nonumber\\
    \!&\!\quad-\eta^{\top}_{2}(\kappa-d(\kappa))Q_{3}\eta_{2}(\kappa-d(\kappa)),\nonumber\\
		 \!&=\! \xi^{\top}(\kappa)\Phi_{2}\xi(\kappa),\label{eq18}\\[+3pt]
	\Delta V_{3}(\kappa)\!&=\!  \eta_{3}^{\top}(\kappa-1)\left[d_{21}R+d_{l}S\right]\eta_{3}(\kappa-1)-\mathrm{J}_{1}\!-\!\mathrm{J}_{2}\!-\!\mathrm{J}_{3},\nonumber\\
        \!&=\!\xi^{\top}(\kappa)\Phi_{3}\xi(\kappa)-\mathrm{J}_{1}-\mathrm{J}_{2}-\mathrm{J}_{3},\label{eq19}\vspace{-5pt}
	\end{align}
    where
     \vspace{-5pt}
    \begin{align*}
        \mathrm{J}_{1}&=\sum_{i=\kappa-d(\kappa)}^{\kappa-d_{l}-1}\eta_{3}^{\top}(i)R\eta_{3}(i),\quad \mathrm{J}_{2}=\sum_{i=\kappa-d_{u}}^{\kappa-d(\kappa)-1}\eta_{3}^{\top}(i)R\eta_{3}(i),\\[-1pt]
        \mathrm{J}_{3}&=\sum_{i=\kappa-d_{l}}^{\kappa-1}\eta_{3}^{\top}(i)S\eta_{3}(i).\vspace{-5pt}
    \end{align*}
Using Lemma~\ref{lem3} in conjunction with \eqref{eq14}, the terms $\mathrm{J}_i$ for $i = 1, 2, 3$ are estimated as follows:
  \vspace{-5pt}
     \begin{align}
    	&\resizebox{0.88\hsize}{!}{$\mathrm{J}_{1}\geq \xi^{\top}(\kappa)\Psi_{11}(d(\kappa))\xi(\kappa)+\frac{1}{d_{\kappa1}}\xi^{\top}(\kappa)F^{\top}_{11}\Psi_{21}F_{11}\xi(\kappa),$}\label{eq20}\\
    	&\resizebox{0.88\hsize}{!}{$\mathrm{J}_{2}\geq \xi^{\top}(\kappa)\Psi_{12}(d(\kappa))\xi(\kappa)+\frac{1}{d_{\kappa2}}\xi^{\top}(\kappa)F^{\top}_{12}\Psi_{22}F_{12}\xi(\kappa)$},\label{eq21}\\
        &\mathrm{J}_{3}\geq \xi^{\top}(\kappa)\Psi_{3}\xi(\kappa),\label{eq22}\vspace{-5pt}
    \end{align}
     where  $\Psi_{11}(d(\kappa)),\Psi_{12}(d(\kappa)),\Psi_{21},\Psi_{22},\Psi_{3}, F_{11}$ and $F_{12}$ are defined in Theorem~\ref{Theorem1}.\\
   Using \eqref{eq16} and \eqref{eq17}–\eqref{eq22}, we arrive at:  
    \vspace{-5pt}
   \begin{align}
    \Delta V(\kappa)&\leq  \;\xi^{\top}(\kappa)\big[\Phi_{1}(d(\kappa))+\Phi_{2}+\Phi_{3}-\Psi_{11}(d(\kappa))\nonumber\\[-1pt]
    &\quad-\Psi_{12}(d(\kappa))-\Psi_{3}
    -\frac{1}{d_{\kappa1}}F^{\top}_{11}\Psi_{21}F_{11}\nonumber\\[-1pt]
    &\quad-\frac{1}{d_{\kappa2}}F^{\top}_{12}\Psi_{22}F_{12}\big]\xi(\kappa).\label{eq23}\vspace{-5pt}
    \end{align}
For any matrices $X_1$ and $X_2$ with appropriate dimensions, by applying Lemma~\ref{lem2} with $\alpha = \frac{d_{\kappa1}}{d_{21}}$ and utilizing \eqref{eq14}, we find:
 \vspace{-3pt}
\begin{align}
    \Delta V(\kappa)&\leq  \;\xi^{\top}(\kappa)\big[\Phi_{1}(d(\kappa))+\Phi_{2}+\Phi_{3}-\Psi_{11}(d(\kappa))\nonumber\\[-1pt]
    &\quad-\Psi_{12}(d(\kappa))-\Psi_{3}-\Psi_{4}(d(\kappa))\big]\xi(\kappa)\nonumber\\[-1pt]
    &\quad+\frac{d_{\kappa1}}{d_{21}^{2}}\xi^{\top}(\kappa)F^{\top}_{12}X^{\top}_{1}\Psi_{21}^{-1}X_{1}F_{12}\xi(\kappa)\nonumber\\[-1pt]
    &\quad+\frac{d_{\kappa2}}{d_{21}^{2}}\xi^{\top}(\kappa)F^{\top}_{11}X_{2}\Psi_{22}^{-1}X^{\top}_{2}F_{11}\xi(\kappa).\label{eq24}\vspace{-5pt}
    \end{align}
    In addition, the definition of $\xi(\kappa)$ implies that several vectors contained within it satisfy the following relations:
    \vspace{-5pt}
     \begin{align*}
	    v_{1}(d(\kappa),d_{l})&=\zeta(\kappa-d_{l})+x(\kappa-d_{l})-x(\kappa-d(\kappa)),\\[-1pt]
        v_{1}(d_{u},d(\kappa))&=\zeta(\kappa-d(\kappa))+x(\kappa-d(\kappa))-x(\kappa-d_{u}),\\[-1pt]
        v_{2}(d(\kappa),d_{l})&=q_{1}(d_{\kappa1})\left[\zeta(\kappa-d_{l})+x(\kappa-d_{l})\right]\\[-1pt]
        &\quad-u_{1}(d(\kappa),d_{l}),\\[-1pt]
        v_{2}(d_{u},d(\kappa))&=q_{1}(d_{\kappa2})\left[\zeta(\kappa-d(\kappa)))+x(\kappa-d(\kappa))\right]\\[-1pt]
        &\quad-u_{1}(d_{u},d(\kappa)).\vspace{-5pt}
     \end{align*}
     These relations result in the following zero-value terms:
       \vspace{-5pt}
        \begin{align}
	    0\!&=\!\xi^{\top}(\kappa)\Phi_{41}(d(\kappa))\xi(\kappa),\!\quad\!
        0\!=\!\xi^{\top}(\kappa)\Phi_{42}(d(\kappa))\xi(\kappa),\label{eq25}\\
        0\!&=\!\xi^{\top}(\kappa)\Phi_{43}(d(\kappa))\xi(\kappa),\!\quad\!
        0\!=\!\xi^{\top}(\kappa)\Phi_{44}(d(\kappa))\xi(\kappa),\label{eq26}\vspace{-5pt}
     \end{align}
     where $G_1$, $G_2$, $G_3$, and $G_4$ are any matrices.\\
    Adding \eqref{eq25} and \eqref{eq26} to \eqref{eq24}, we obtain that
    \vspace{-5pt}
    \begin{align}\label{eq29}
    \Delta V(\kappa)&\leq \xi^{\top}(\kappa)\big[\Xi(d(\kappa))+\frac{d_{\kappa1}}{d_{21}^{2}}F^{\top}_{12}X^{\top}_{1}\Psi_{21}^{-1}X_{1}F_{12}\nonumber\\[-1pt]
    &\quad+\frac{d_{\kappa2}}{d_{21}^{2}}F^{\top}_{11}X_{2}\Psi_{22}^{-1}X^{\top}_{2}F_{11}\big]\xi(\kappa),\vspace{-5pt}
    \end{align}
   where $\Xi(d(\kappa))$ is defined in Theorem~~\ref{Theorem1}. Applying the Schur complement to~\eqref{eq15} shows that the right-hand 
side of~\eqref{eq29} is negative definite. Since $\Xi(d(\kappa))$ is affine 
in $d(\kappa)$, convexity ensures that~\eqref{eq15} holding at the two 
vertices $d(\kappa)\in\{d_l,d_u\}$ implies $\Delta V(\kappa)<0$ for all 
nonzero $\xi(\kappa)$ and all $d(\kappa)\in[d_l,d_u]$. Combined with 
$V(\kappa)>0$, Lyapunov stability theory guarantees asymptotic stability 
of~\eqref{eq2}. This completes the proof.
\end{proof}

\begin{remark}\label{rem3}
The reduced conservatism of Theorem~\ref{Theorem1} relative to
\cite{ge2022delay,zhang2022novel,zhang2023stabilitya,zhang2023stabilityb,%
wang2024delay,xie2024stability,luo2025new,chen2020stability} stands on three pillars. First, the augmented LKF~\eqref{eq16} enriches
the standard state information with $x(\kappa-1)$, the boundary terms
$\zeta(\kappa-d_{l})$, $\zeta(\kappa-d(\kappa))$, $\zeta(\kappa-d_{u})$,
and the weighted summations $v_{1}(d(\kappa),d_{l}), v_{1}(d_{u},d(\kappa)), v_{2}(d(\kappa),d_{l}), v_{2}(d_{u},d(\kappa))$, capturing richer delay-dependent
and higher-order difference information. Second, the IMSBI
(Lemma~\ref{lem3}) yields tighter bounds on $J_1$, $J_2$, and $J_3$ by
explicitly exploiting the coupling among $x(i)$, $\zeta(i)$, and
$\Delta\zeta(i)$. Third, the zero-equations~\eqref{eq25}--\eqref{eq26} and free matrices $G_i$,
$i=1,\dots,4$, inject additional cross-terms that expand the feasibility
region of the LMIs.
\end{remark}
\section{Numerical examples}\label{sec4}
This section validates Theorem~\ref{Theorem1} via two numerical examples. The MATLAB code is available at Github\footnote{\url{https://github.com/Echakroune/Th1_MATLAB_Code.git}}. We assess conservatism via the maximal admissible upper bounds (MAUBs) and computational complexity via the number of decision variables (NDVs).
\begin{example}\label{ex1}
Consider the DTS \eqref{eq2} with  matrices given by:
	\begin{equation*}
    B=\left[\begin{array}{cc} 1.00 & 0.01 \\ -0.10 & 0.99 \end{array}\right],\quad B_{d}=\left[\begin{array}{cc} 0.003 & 0.001 \\ 0.010 & 0.005 \end{array}\right].\vspace{-5pt}
    \end{equation*}
Assuming $\Delta d(\kappa)\triangleq d(\kappa+1)-d(\kappa)\in[-\mu,\mu]$, we compute the MAUBs $d_{u}$ for various $d_{l}$ and $\mu$. Results from Theorem~\ref{Theorem1} (Table~\ref{tab1}) are compared with several recent methods. The proposed approach yields larger $d_{u}$ values than those in \cite{ge2022delay,zhang2023stabilityb,wang2024delay,xie2024stability,luo2025new,chen2020stability}, indicating reduced conservatism due to the augmented LKF, the IMSBI, and the zero equations \eqref{eq25}--\eqref{eq26}. Although the NDVs here exceed those in \cite{ge2022delay,zhang2023stabilityb,chen2020stability}, the computational cost remains lower than in \cite{wang2024delay,xie2024stability,luo2025new}.
\begin{table}[ht!]
\centering
\vspace{-3pt}
\caption{The MAUBs $d_{u}$ and NDVs for varying $d_{l}$ and $\mu$ (Example~\ref{ex1}).}
\label{tab1}
\footnotesize
\setlength{\tabcolsep}{4pt}
\begin{tabular}{l c c c c c c c}
\toprule
& \multicolumn{6}{c}{$d_{l}$} \\
\cmidrule(lr){3-7}
Criteria & $\mu$ & 15 & 25 & 35 & 45 & 50 & NDVs \\
\midrule
\text{Th.}1~\cite{chen2020stability} & -- & 120 & 135 & 146 & 156 & 160 & $39.5n^2+6.5n$ \\
\text{Th.}1~\cite{zhang2023stabilityb}  & -- & 135 & 143 & 154 & 166 & 171 & $75.5n^2+8.5n$ \\
\midrule
\multirow{2}{*}{\text{Th.}3.1~\cite{ge2022delay}} & 1 & 123 & 145 & 157 & 167 & 169 & \multirow{2}{*}{$40.5n^2+7.5n$} \\
                              & 5 & 120 & 136 & 146 & 159 & 163 & \\
\midrule
\multirow{2}{*}{\text{Th.}1~\cite{wang2024delay}} & 1 & 138 & 159 & 163 & 166 & 167 & \multirow{2}{*}{$372n^2+7n$} \\
                              & 5 & 128 & 142 & 151 & 158 & 162 & \\
\midrule
\multirow{2}{*}{\text{Th.}2~\cite{wang2024delay}} & 1 & 141 & 162 & 166 & 168 & 169 & \multirow{2}{*}{$372n^2+7n$} \\
                              & 5 & 138 & 154 & 162 & 167 & 168 & \\
\midrule
\multirow{2}{*}{\text{Th.}2~\cite{zhang2023stabilityb}} & 1 & 153 & 165 & 170 & 173 & 175 & \multirow{2}{*}{$82n^2+12n$} \\
                                    & 5 & 139 & 149 & 157 & 167 & 170 & \\
\midrule
\multirow{2}{*}{\text{Th.}1.C1~\cite{xie2024stability}} & 1 & 136 & 156 & 168 & 173 & 177 & \multirow{2}{*}{$267.5n^2+13.5n$} \\
                                     & 5 & 128 & 149 & 161 & 169 & 171 & \\
\midrule
\multirow{2}{*}{\text{Th.}1.C2~\cite{xie2024stability}~} & 1 & 150 & 165 & 172 & 176 & 178 & \multirow{2}{*}{$267.5n^2+13.5n$} \\
                                     & 5 & 147 & 162 & 170 & 173 & 176 & \\
\midrule
\multirow{2}{*}{\text{Th.}1~\cite{luo2025new}~} & 1 & 154 & 166 & 171 & 174 & 175 & \multirow{2}{*}{$420n^2+12n$} \\
                            & 5 & 136 & 152 & 160 & 167 & 169 & \\
\midrule
Theorem~\ref{Theorem1} & -- & 180 & 207 & 249 & 282 & 290 & $187n^2+16n$ \\
\bottomrule
\end{tabular}
\end{table}
To further validate the stability criterion proposed in Theorem~\ref{Theorem1}, a numerical simulation is conducted for Example~\ref{ex1}. The DTS described by \eqref{eq2} is examined under a TVD randomly selected within the interval $[25,\ 207]$, with initial conditions $\varphi(\kappa) = [0.8,\ -0.5]^T$ for $\kappa \in [-207,\ 0]$. As shown in Figure~\ref{fig1}, the state trajectories converge asymptotically to zero, thereby confirming the system’s stability and supporting the theoretical findings.
\begin{figure}[ht!]
    \centering
    \vspace{-5pt}
    \includegraphics[width=\linewidth]{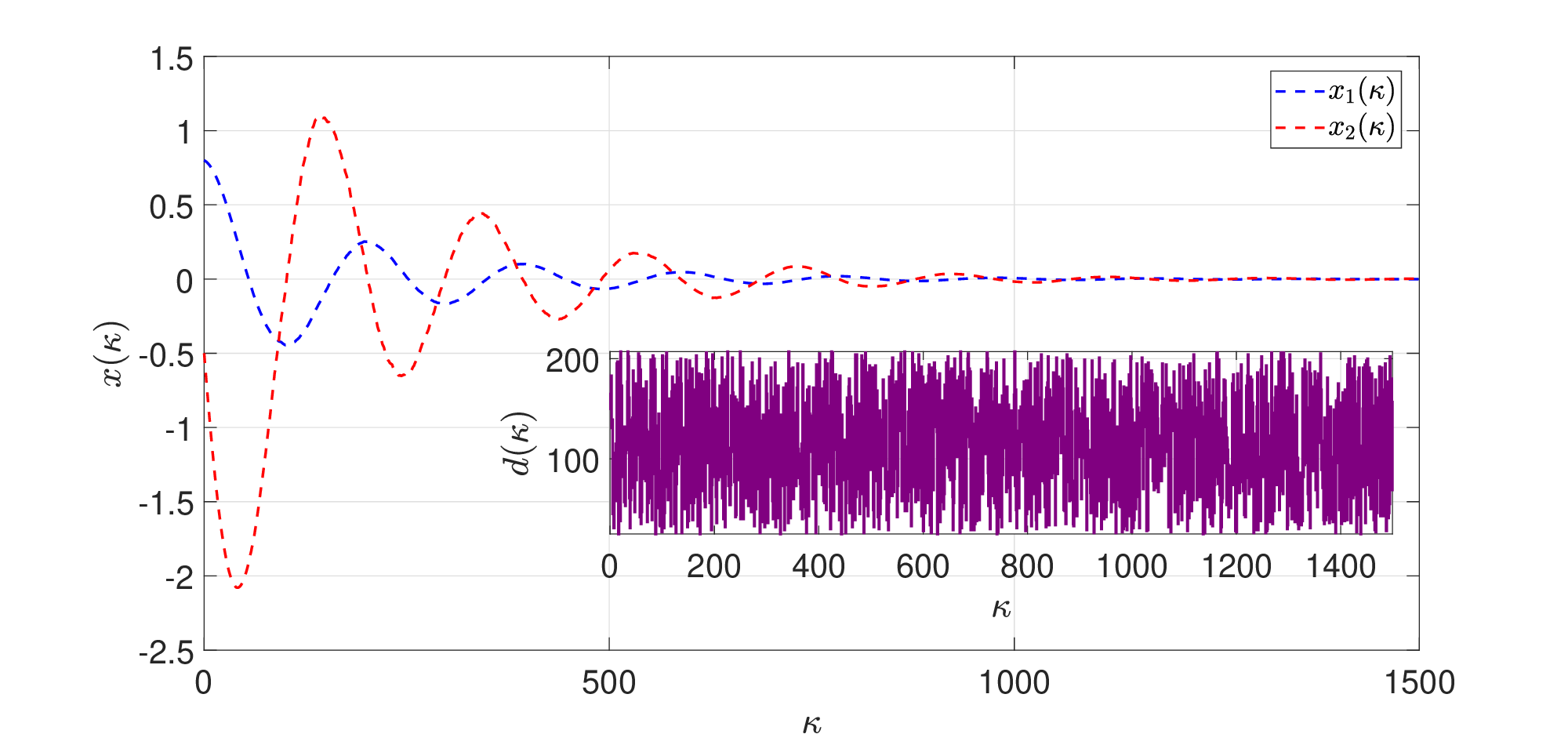}\vspace{-10pt}
   \caption{State trajectories and corresponding time delays in Example~\ref{ex1}.}
    \label{fig1}\vspace{-5pt}
\end{figure}
\end{example}
\begin{example}\label{ex2}
Consider the DTS \eqref{eq2} with  matrices given by:
\vspace{-5pt}
	\begin{equation*}
    B=\left[\begin{array}{cc} 0.80 & 0.00 \\ 0.05 & 0.90 \end{array}\right],\quad B_{d}=\left[\begin{array}{cc} -0.1 & 0.0 \\ -0.2 & -0.1 \end{array}\right].\vspace{-5pt}
    \end{equation*}
\begin{table}[ht!]
\centering
\caption{The MAUBs $d_{u}$ and NDVs for varying $d_{l}$ (Example~\ref{ex2}).}
\label{tab2}
\footnotesize
\setlength{\tabcolsep}{4pt}
\begin{tabular}{l c c c c c c c}
\toprule
& \multicolumn{6}{c}{$d_{l}$} \\
\cmidrule(lr){2-7}
Criteria & 2 & 6 & 10 & 15 & 20 & 25 & NDVs \\
\midrule
\text{Th.}1~\cite{zhang2023stabilityb}~ & 22 & 22 & 23 & 25 & 29 & 32 & $75.5n^2+8.5n$ \\
\text{Th.}1~\cite{chen2020stability} & 22 & 22 & 23 & 26 & 29 & 32 & $39.5n^2+6.5n$ \\
\text{Cor.}8~\cite{zhang2022novel} & 22 & 23 & 23 & 26 & 29 & 32 & $1280.5n^2+13.5n$ \\
\text{Th.}3.1~\cite{ge2022delay} & 22 & 23 & 23 & 26 & 29 & 32 & $40.5n^2+7.5n$ \\
\text{Th.}2~\cite{zhang2023stabilityb} & 22 & 23 & 24 & 26 & 29 & 33 & $82n^2+12n$ \\
\text{Prop.}3~\cite{zhang2023stabilitya} & 23 & 24 & 25 & 27 & 30 & 32 & $341.5n^2+8.5n$ \\
\text{Th.}1~\cite{luo2025new} & 23 & 25 & 26 & 28 & 31 & 34 & $420n^2+12n$ \\
\text{Th.}1~\cite{wang2024delay} & 23 & 25 & 26 & 28 & 31 & 34 & $372n^2+7n$ \\
\text{Th.}2~\cite{wang2024delay} & 24 & 25 & 28 & 31 & 34 & 36 & $372n^2+7n$ \\
\text{Th.}1.C1~\cite{xie2024stability} & 23 & 25 & 27 & 30 & 33 & 37 & $267.5n^2+13.5n$ \\
\text{Th.}1.C2~\cite{xie2024stability} & 23 & 25 & 27 & 30 & 34 & 38 & $267.5n^2+13.5n$ \\
\midrule
Theorem~\ref{Theorem1} & 37 & 38 & 39 & 41 & 43 & 47 & $187n^2+16n$ \\
\bottomrule
\end{tabular}
\end{table}
Table~\ref{tab2} compares the MAUBs $d_{u}$ (for various $d_{l}$) and the corresponding NDVs obtained from Theorem~\ref{Theorem1} and several existing methods \cite{ge2022delay,zhang2022novel,zhang2023stabilitya,zhang2023stabilityb,wang2024delay,xie2024stability,luo2025new,chen2020stability}. The comparison assumes $\Delta d(\kappa)\in[-\mu,\mu]$ with $\mu=1$ for \cite{ge2022delay,zhang2023stabilitya,zhang2023stabilityb,wang2024delay,xie2024stability,luo2025new}. Theorem~\ref{Theorem1} consistently yields larger MAUBs while requiring relatively few NDVs, indicating reduced conservatism and improved computational efficiency.
\begin{figure}[ht!]
    \centering
    \vspace{-5pt}
    \includegraphics[width=\linewidth]{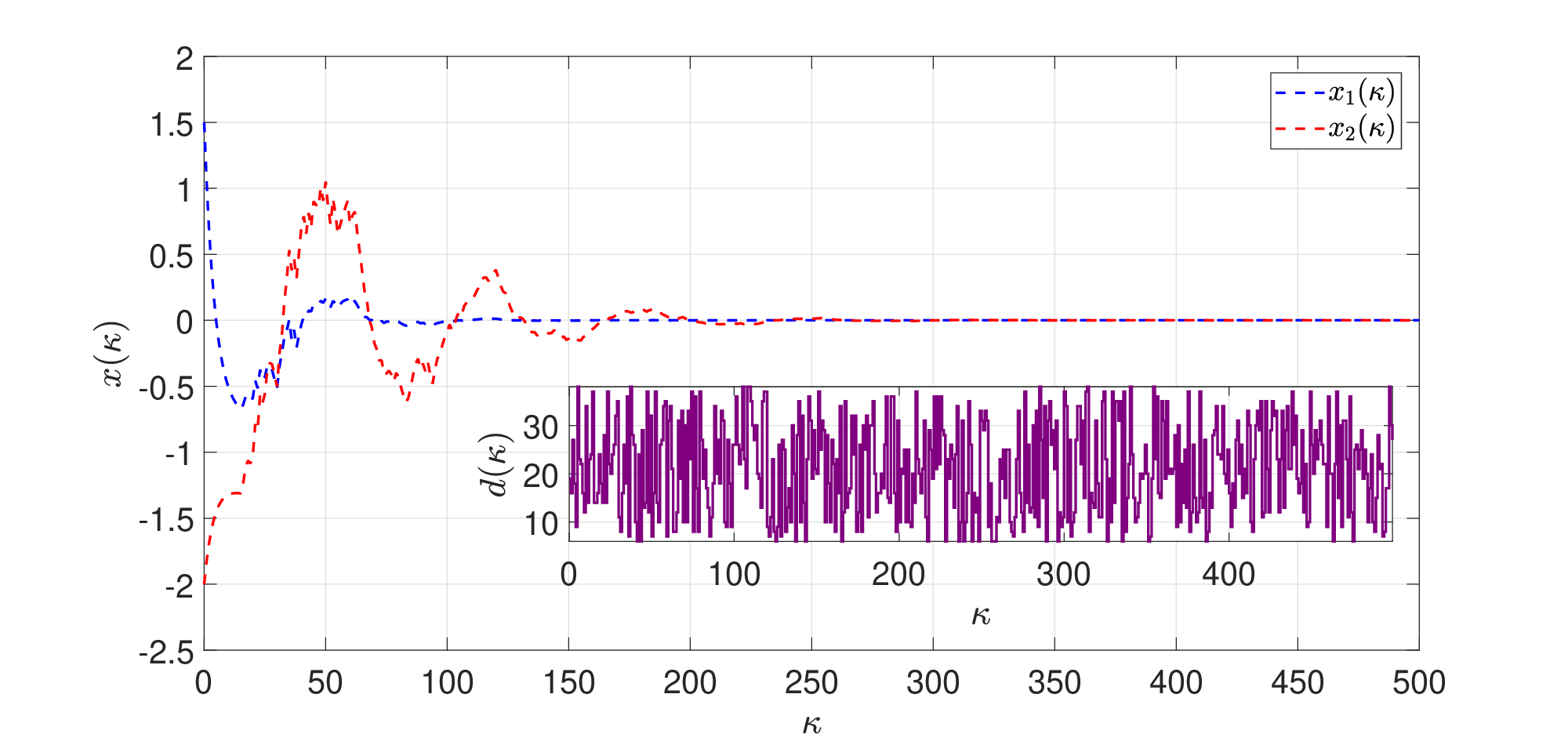}\vspace{-10pt}
   \caption{State trajectories and corresponding time delays in Example~\ref{ex2}.}
    \label{fig2}\vspace{-15pt}
\end{figure}
Example~\ref{ex2} numerically verifies Theorem~\ref{Theorem1} for system \eqref{eq2} using stochastically generated delays in $[6,38]$ and initial states $\varphi(\kappa)=[1.5,\,-2.0]^\top$ for $\kappa\in[-38,0]$. As shown in Figure~\ref{fig2}, both states converge asymptotically to the origin, confirming the theoretical result.
\end{example}	
\section{Conclusion}
This letter proposed a less conservative framework for stability analysis of discrete-time systems with time-varying delays. The main contributions are an improved matrix-separation-based inequality for tighter estimation of augmented summation terms and an augmented LKF that captures state–delay interactions. Embedding these elements into the forward-difference estimate yields new delay-dependent stability conditions cast as tractable LMIs. Two numerical examples confirm that the approach consistently produces less conservative bounds while retaining computational efficiency.

\vspace{-10pt}

\bibliographystyle{IEEEtran}
\bibliography{biblio}

@article{van2012discrete,
  title={A discrete-time framework for stability analysis of nonlinear networked control systems},
  author={van de Wouw, Nathan and Ne{\v{s}}i{\'c}, D and Heemels, WPMH},
  journal={Automatica},
  volume={48},
  number={6},
  pages={1144--1153},
  year={2012},
  publisher={Elsevier}
}

@article{yan2022lyapunov,
  title={Lyapunov-function-based event-triggered control of nonlinear discrete-time cyber--physical systems},
  author={Yan, Shen and Gu, Zhou and Park, Ju H},
  journal={IEEE Transactions on Circuits and Systems II: Express Briefs},
  volume={69},
  number={6},
  pages={2817--2821},
  year={2022},
  publisher={IEEE}
}

@article{zadeh2016discrete,
  title={Discrete-time modeling, stability analysis, and active stabilization of DC distribution systems with multiple constant power loads},
  author={Zadeh, Mehdi Karbalaye and Gavagsaz-Ghoachani, Roghayeh and Martin, Jean-Philippe and Nahid-Mobarakeh, Babak and Pierfederici, Serge and Molinas, Marta},
  journal={IEEE Transactions on Industry Applications},
  volume={52},
  number={6},
  pages={4888--4898},
  year={2016},
  publisher={IEEE}
}

@book{gu2003stability,
  title={Stability of time-delay systems},
  author={Gu, Keqin and Chen, Jie and Kharitonov, Vladimir L},
  year={2003},
  publisher={Springer Science \& Business Media}
}

@article{seuret2015hierarchy,
  title={Hierarchy of LMI conditions for the stability analysis of time-delay systems},
  author={Seuret, Alexandre and Gouaisbaut, Fr{\'e}d{\'e}ric},
  journal={Systems \& Control Letters},
  volume={81},
  pages={1--7},
  year={2015},
  publisher={Elsevier}
}

@article{kwon2013stability,
  title={Stability and stabilization for discrete-time systems with time-varying delays via augmented Lyapunov--Krasovskii functional},
  author={Kwon, OM and Park, Myeong-Jin and Park, Ju H and Lee, Sang-Moon and Cha, Eun-Jong},
  journal={Journal of the Franklin Institute},
  volume={350},
  number={3},
  pages={521--540},
  year={2013},
  publisher={Elsevier}
}

@inproceedings{jiang2005stability,
  title={Stability criteria for linear discrete-time systems with interval-like time-varying delay},
  author={Jiang, Xiefu and Han, Qing-Long and Yu, Xinghuo},
  booktitle={Proceedings of the 2005, American Control Conference, 2005.},
  pages={2817--2822},
  year={2005},
  organization={IEEE}
}

@article{nam2015discretea,
  title={Discrete Wirtinger-based inequality and its application},
  author={Nam, Phan T and Pathirana, Pubudu N and Trinh, Hieu},
  journal={Journal of the Franklin Institute},
  volume={352},
  number={5},
  pages={1893--1905},
  year={2015},
  publisher={Elsevier}
}

@article{nam2015discreteb,
  title={Discrete inequalities based on multiple auxiliary functions and their applications to stability analysis of time-delay systems},
  author={Nam, Phan Thanh and Trinh, Hieu and Pathirana, Pubudu N},
  journal={Journal of the Franklin Institute},
  volume={352},
  number={12},
  pages={5810--5831},
  year={2015},
  publisher={Elsevier}
}

@article{park2011reciprocally,
  title={Reciprocally convex approach to stability of systems with time-varying delays},
  author={Park, PooGyeon and Ko, Jeong Wan and Jeong, Changki},
  journal={Automatica},
  volume={47},
  number={1},
  pages={235--238},
  year={2011},
  publisher={Elsevier}
}

@article{nabil2024improved,
  title={Improved delay-dependent stability analysis of digital filters with generalized overflow arithmetic},
  author={Echakroune, Nabil and Tissir, El Houssaine and  Hmamed, Abdelaziz},
  journal={Journal of the Franklin Institute},
  volume={361},
  number={16},
  pages={107157},
  year={2024},
  publisher={Elsevier}
}

@article{park2016stability,
  title={Stability and stabilization of discrete-time {T--S} fuzzy systems with time-varying delay via Cauchy--Schwartz-based summation inequality},
  author={Park, Myeong-Jin and Kwon, Oh-Min},
  journal={IEEE Transactions on Fuzzy Systems},
  volume={25},
  number={1},
  pages={128--140},
  year={2016},
  publisher={IEEE}
}

@article{park2017stability,
  title={Stability analysis of discrete-time switched systems with time-varying delays via a new summation inequality},
  author={Park, MJ and Kwon, OM and Choi, SG},
  journal={Nonlinear Analysis: Hybrid Systems},
  volume={23},
  pages={76--90},
  year={2017},
  publisher={Elsevier}
}

@article{chen2022delay,
  title={Delay-variation-dependent criteria on stability and stabilization for discrete-time {T--S} fuzzy systems with time-varying delays},
  author={Chen, Wen-Hu and Zhang, Chuan-Ke and Xie, Ke-You and Zhu, Cui and He, Yong},
  journal={IEEE Transactions on Fuzzy Systems},
  volume={30},
  number={11},
  pages={4856--4866},
  year={2022},
  publisher={IEEE}
}

@article{sha2024stability,
  title={Stability analysis of discrete-time systems with a time-varying delay via improved methods},
  author={Sha, Hongjia and Park, Ju H and Chen, Jun and Zhu, Mingbo and Nan, Chengjie},
  journal={IET Control Theory \& Applications},
  volume={18},
  number={7},
  pages={951--959},
  year={2024},
  publisher={Wiley Online Library}
}

@article{zhang2017stability,
  title={Stability analysis of discrete-time neural networks with time-varying delay via an extended reciprocally convex matrix inequality},
  author={Zhang, Chuan-Ke and He, Yong and Jiang, Lin and Wang, Qing-Guo and Wu, Min},
  journal={IEEE Transactions on Cybernetics},
  volume={47},
  number={10},
  pages={3040--3049},
  year={2017},
  publisher={IEEE}
}

@article{chen2020stability,
  title={Stability of discrete-time systems with time-varying delay via a novel Lyapunov-Krasovskii functional},
  author={Chen, Jun and Park, Ju H and Xu, Shengyuan and Zhang, Xian-Ming},
  journal={International Journal of Robust and Nonlinear Control},
  volume={30},
  number={12},
  pages={4779--4788},
  year={2020},
  publisher={Wiley Online Library}
}

@article{ge2022delay,
  title={A delay-variation-dependent stability criterion for discrete-time systems via a bivariate quadratic function negative-determination lemma},
  author={Ge, Xiao and Hoi, Kachon and Vong, Seakweng},
  journal={Journal of the Franklin Institute},
  volume={359},
  number={10},
  pages={4976--4996},
  year={2022},
  publisher={Elsevier}
}

@article{zhang2022novel,
  title={A novel approach to {$H_\infty$} performance analysis of discrete-time networked systems subject to network-induced delays and malicious packet dropouts},
  author={Zhang, Xian-Ming and Han, Qing-Long and Ge, Xiaohua},
  journal={Automatica},
  volume={136},
  pages={110010},
  year={2022},
  publisher={Elsevier}
}

@article{zhang2023stabilitya,
  title={Stability analysis of delayed discrete-time systems based on a delay-square-dependent Lyapunov functional},
  author={Zhang, Xian-Ming and Han, Qing-Long and Ge, Xiaohua and Peng, Chen},
  journal={Automatica},
  volume={147},
  pages={110592},
  year={2023},
  publisher={Elsevier}
}

@article{zhang2023stabilityb,
  title={Stability analysis of discrete-time systems with time-varying delay via a delay-dependent matrix-separation-based inequality},
  author={Zhang, Chuan-Ke and Chen, Wen-Hu and Zhu, Cui and He, Yong and Wu, Min},
  journal={Automatica},
  volume={156},
  pages={111192},
  year={2023},
  publisher={Elsevier}
}

@article{wang2024delay,
  title={Delay-variation-dependent summation inequality and its application to stability analysis of discrete-time systems with time-varying delay},
  author={Wang, Chen-Rui and He, Yong and Zhang, Chuan-Ke and Chen, Wen-Hu and Wu, Min},
  journal={Systems \& Control Letters},
  volume={184},
  pages={105721},
  year={2024},
  publisher={Elsevier}
}

@article{xie2024stability,
  title={Stability analysis of discrete-time delayed systems via matrix-injection-based transformation method for bivariate quadratic functions},
  author={Xie, Ke-You and Shangguan, Xing-Chen and Wang, Hong-Zhang and Jin, Li},
  journal={Journal of the Franklin Institute},
  volume={361},
  number={6},
  pages={106757},
  year={2024},
  publisher={Elsevier}
}

@article{luo2025new,
  title={New stability criteria for discrete-time systems with a time-varying delay via a delay-variation-dependent free-matrix-based inequality},
  author={Luo, Li and Chen, Jun and Song, Bo and Zhuang, Guangming},
  journal={International Journal of Dynamics and Control},
  volume={13},
  number={4},
  pages={131},
  year={2025},
  publisher={Springer}
}
\end{document}